\documentclass[usenatbib,useAMS]{mn2e} 
\pdfoutput=1
\newcommand{\hilite}{}

\newcommand{\glite}{}

\usepackage{graphicx}
\usepackage{amsmath}
\usepackage[usenames,dvipsnames]{color}

\newcommand{\half}{\frac{1}{2}}
\newcommand{\quart}{\frac{1}{4}}

\newcommand{\rd}{\rmn{d}}

\newcommand{\pderiv}[2]{{\upartial#1\over\upartial#2}}

\newcommand{\deriv}[2]{{\rd#1\over \rd#2}}

\newcommand{\D}{{\mathbf{D}}}
\newcommand{\X}{{\mathbf{X}}}

\renewcommand{\a}{{\bmath{a}}}
\renewcommand{\b}{{{\bmath{b}}}}
\newcommand{\e}{{\hat{\bmath{e}}}}
\renewcommand{\j}{{\bmath{j}}}
\renewcommand{\k}{{\bmath{k}}}
\newcommand{\m}{{\bmath{m}}}
\newcommand{\F}{{{\bmath{F}}}}
\newcommand{\g}{{{\bmath{g}}}}
\newcommand{\irom}{{\rmn{i}}}

\newcommand{\calL}{{\mathcal{L}}}
\newcommand{\calD}{{\mathcal{D}}}

\renewcommand{\v}{{\bmath{v}}}
\newcommand{\x}{{\bmath{x}}}
\newcommand{\vdot}{{\bmath{\cdot}}}
\newcommand{\grad}{\bmath{\nabla}}
\newcommand{\B}{{\bmath{B}}}
\newcommand{\E}{{\bmath{E}}}
\newcommand{\vcross}{{\bmath{\times}}}

\newcommand{\thth}{\hspace{1.5pt}}
\newcommand{\curl}{\grad\vcross}
\newcommand{\Curl}{\grad\vcross\thth}

\newcommand\Div{\grad\vdot\thth}

\newcommand{\kpar}{k_{\scriptscriptstyle\parallel}}

\newcommand{\bv}{Brunt-V\"ais\"al\"a}

\renewcommand{\Re}{\mathop{\rm Re}\nolimits}

\newcommand{\mnras}{MNRAS}
\newcommand{\apj}{ApJ}
\newcommand{\apjl}{ApJL}
\newcommand{\apjs}{ApJS}
\newcommand{\aap}{A\&A}
\newcommand{\solphys}{Solar Phys.}

\begin{document}

%\singlespace

\title[Magneto-Gravity Waves in the Solar Chromosphere]{Reflection and Conversion of Magneto-Gravity Waves in the Solar Chromosphere:\\ Windows to the Upper Atmosphere}

\author[Marie Newington and Paul Cally]{
Marie E. Newington\thanks{E-mail: marie.gibbon@sci.monash.edu.au} and
Paul S. Cally\thanks{E-mail: paul.cally@sci.monash.edu.au}\\
Centre for Stellar and Planetary Astrophysics,
School of Mathematical Sciences,
Monash University, Victoria, Australia 3800}

\maketitle

\begin{abstract}
The detection of upward propagating internal gravity waves {\glite at the base of} the Sun's chromosphere has recently been reported by Straus et al., who postulated that these may efficiently couple to Alfv\'en waves in magnetic regions. This may be important in transporting energy to higher levels. Here we explore the propagation, reflection and mode conversion of linear gravity waves in a VAL C atmosphere, and find that even weak magnetic fields usually reflect gravity waves back downward as slow magnetoacoustic waves well before they reach the Alfv\'en/acoustic equipartition height at which mode conversion might occur. However, for certain highly inclined magnetic field orientations in which the gravity waves manage to penetrate near or through the equipartition level, there can be substantial conversion to either or both upgoing Alfv\'en and acoustic waves. Wave energy fluxes comparable to the chromospheric radiative losses are expected.
\end{abstract}

%%%%%%%%%%%%%%%%%%%%%%%%%%%%%%%%%%%%%%%%%%%%%%%%%%%%%
\section{INTRODUCTION}                \label{intro}
Using the Interferometric BIdimensional Spectrometer (IBIS) and the Echelle Spectrograph on the Dunn Solar Telescope (DST) of the Sacramento Peak National Solar Observatory, and the Michelson Doppler Imager (MDI) on the Solar and Heliospheric Observatory (SOHO), \cite{straus08} have identified upward propagating\footnote{The group velocity and hence energy flux is upward. As expected of gravity waves, the phase velocity is downward.}  gravity waves with frequencies between 0.7 mHz and 2.1 mHz in weak magnetic field regions of the solar atmosphere, and found that their energy flux is an order of magnitude larger than co-spatial acoustic waves, and comparable to the expected quiet-sun chromospheric losses of around 4.3 $\rm kW\,m^{-2}$. However, the gravity waves were found to be significantly suppressed in stronger field regions. Nevertheless, in light of recent identification of ubiquitous Alfv\'en waves in the corona \citep{pontieu07,tomczyk07}, they postulate that when these gravity waves enter magnetic regions they may efficiently couple to Alfv\'en waves, perhaps contributing to the observed coronal wave flux. Furthermore, \cite{jess09} have directly identified torsional Alfv\'en waves in H$\alpha$ bright-point groups at frequencies as low as 1.4 mHz, which places them at least partially in the gravity wave regime if there is any coupling.

In this paper, we explore the propagation, reflection, and mode conversion of atmospheric gravity waves of around 1 mHz in frequency using both dispersion relations and numerical solution of the governing differential equations in simple atmospheric models with uniform inclined magnetic field. The imposed fields of 10 to 100 Gauss are weak in the photospheric context, but become dominant at greater heights as the plasma $\beta$ (the ratio of plasma to magnetic pressure) falls below unity due to density stratification. We therefore have a situation where low frequency waves are essentially gravity waves at low altitudes, but become magnetically dominated at higher levels in the chromosphere. The central question is: \emph{What happens to upward propagating gravity waves as they enter regions where magnetic forces become significant? Does the magnetic field help or hinder propagation through the chromosphere?} The answer is: both, depending on magnetic field orientation.

%%%%%%%%%%%%%%%%%%%%%%%%%%%%%%%%%%%%%%%%%%%%%%%%%%%%%
\section{MODEL AND EQUATIONS}   \label{model}
We adopt the horizontally invariant VAL C Model of \cite{VAL}, as adapted by \cite{sf03}, up to $z=1.6$ Mm. The interesting wave reflections and conversions all happen below this height. An isothermal top is appended above 1.6 Mm. No transition region is included, to avoid the complication of reflections outside our region of interest. 

The linear adiabatic wave equations for this scenario are set out in \cite{cg08}, where exact series solutions are derived for the isothermal case. These are used here to specify top boundary conditions, which are always applied in the isothermal layers. 

{\hilite
Without loss of generality, the waves we consider are assumed to propagate in the $x$-$z$ plane. The orientation of the magnetic field is then adjusted to explore various geometries.} We distinguish the two-dimensional (2D) and three-dimensional (3D) cases:
\begin{description}
\item[2D:] The magnetic field lies in the same vertical ($x$-$z$) plane as the direction of wave propagation, and is inclined an angle $\theta$ from the vertical;
\item[3D:] The vertical plane containing magnetic field lines makes an angle $\phi\ne0$ to the $x$-$z$ plane.
\end{description}
In general, in cartesian coordinates $(x,y,z)$, the magnetic field vector is $\B=B\,(\sin\theta\cos\phi,\, \sin\theta\sin\phi,\,\cos\theta)$, where $B=|\B|$. {\hilite In light of the recent identification of ubiquitous horizontal magnetic field of up to 50--100 G in quiet internetwork regions using the Solar Optical Telescope/Spectro-Polarimeter (SOT/SP) on \emph{Hinode} \citep{lites08}, particular attention shall be given to highly inclined field.}

We use two mathematical tools: the magneto-acoustic-gravity dispersion relation {\hilite (and the ray calculations which derive from it)}, and numerical integration of the $4^{\rm th}$ order (2D) or $6^{\rm th}$ order (3D) wave equations.

%%%%%
\subsection{Dispersion Relation}   \label{sub:disp}
Dispersion relations are widely used to describe oscillations in weakly inhomogeneous media, where the properties of the medium vary slowly on the length scale of a typical wavelength. This property allows us to effectively Fourier analyse in space (and time assuming a steady or slowly varying background) to obtain a relationship between frequency $\omega$ and wavevector $\k$.

The development of dispersion relations is intimately connected with determination of where the solutions of wave equations will be oscillatory and where non-oscillatory. This is a difficult enough task for gravitationally stratified unmagnetized atmospheres \citep{sf03}, let alone magneto-atmospheres. The difficulty ultimately revolves around the definition of the acoustic cutoff frequency $ \omega_{\rm c}$ which, depending on choice of dependent and independent variables, can take many distinct forms. There is simply no such thing as \emph{the} acoustic cutoff frequency, and hence no firm and unambiguous way to precisely distinguish oscillatory and non-oscillatory regions. The most commonly quoted expression is that of \cite{dg84}, {\hilite
\begin{equation}
 \omega_{\rm c}^2= \omega_{\rm\scriptscriptstyle{DG}}^2=\frac{c^2}{4H^2}\left(1-2\deriv{H}{z}\right)\, , \label{cutoffG}
\end{equation}
}where $c$ is the sound speed and $H$ is the density scale height. {\hilite Height $z$ increases upward.} However, as pointed out by \citeauthor{sf03}, the appearance of the second derivative of the density in this expression renders it largely impractical for use with tabulated atmospheres in many cases. With the widely used Model S \citep{S}, it yields an enormous but very thin spike just below the solar surface \citep[see fig.~1 of][]{sc06}, which is inconsistent with the assumption of slow variation of coefficients for any solar waves of practical interest.

Although \cite{sf03} derive more attractive alternatives in the hydrodynamic case, these have not yet been extended to magnetohydrodynamics (MHD). We are therefore in uncertain territory applying dispersion relations to solar atmospheric waves, especially in the few hundred kilometres below the surface. However, above the surface, the situation is less troublesome. All suggested formulations for $ \omega_{\rm c}$ agree in the isothermal atmosphere case, where {\hilite $dH/dz=0$}, returning the so-called isothermal acoustic cutoff frequency
\begin{equation}
 \omega_{\rm c} = \omega_{\rm ci} = \frac{c}{2H}\, . \label{cutoffi}
\end{equation}
Indeed, for an isothermal atmosphere, there are exact solutions which confirm precisely this expression: {\hilite see \citet[\S309]{lamb} for the hydrodynamic case and \cite{cally01} with uniform magnetic field added}. Consequently, in the lower solar atmosphere, where $H$ varies (comparatively) little and slowly, we may adopt equation (\ref{cutoffi}) with the expectation of obtaining at least qualitatively correct results. Furthermore, for the gravity waves which interest us most here, it is the {\bv} frequency $N$ (about which there is no dispute) that is most relevant. 

\begin{figure}
\begin{center}
\includegraphics[width=0.799\hsize]{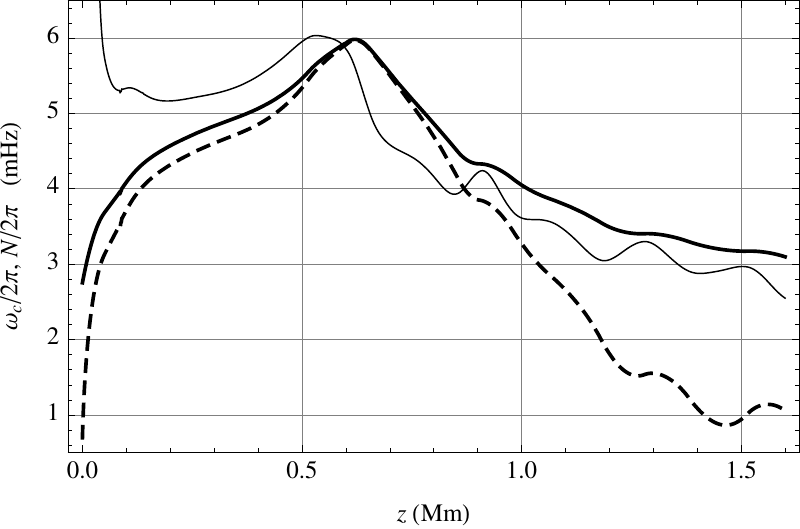}
\caption{Isothermal acoustic cutoff $\omega_{ci}$ (thick full) and {\bv} (dashed) frequencies as functions of height $z$ for the VAL C model atmosphere. The \citeauthor{dg84} cutoff frequency $\omega_{\scriptscriptstyle{DG}}$ (thin full) is shown for comparison.}
\label{fig:freqs}
\end{center}
\end{figure}

Figure \ref{fig:freqs} plots the acoustic cutoff and buoyancy frequencies in the model atmosphere. Most significantly, the {\bv} frequency in the VAL C atmosphere decreases steeply from around 6 mHz to approximately 1 mHz between 0.6 Mm and 1.5 Mm, indicating a propensity to trap gravity waves in the low chromosphere.

The 3D MHD dispersion function 
\begin{multline}
\calD=
\omega^2 \omega_{\rm c}^2 a_y^2  k_{\rm h}^2 +(\omega^2-a^2\kpar^2)\times{}\\
\left[\omega^4-(a^2+c^2)\omega^2 k^2+a^2c^2k^2\kpar^2 \right. \\
\left. \qquad\qquad{}+c^2N^2 k_{\rm h}^2
-(\omega^2-a_z^2k^2) \omega_{\rm c}^2\right],   \label{DR}
\end{multline}
is derived in Appendix \ref{A}. Here an $\exp[\irom(\k\vdot\x-\omega\,t)]$ dependence on position $\x$ and time $t$ is assumed, $ k_{\rm h}=k_x$ and $\kpar$ are respectively the horizontal and field-aligned components of the wavevector, $k=|\k|$ is the wavenumber, $a$ is the Alfv\'en speed, $a_y$ is the component of the Alfv\'en velocity in the $y$-direction (perpendicular to the vertical plane of propagation) and $a_z$ its vertical component, $N$ is the {\bv} (buoyancy) frequency given by 
{\hilite 
\begin{equation}
N^2=\frac{g}{H}-\frac{g^2}{c^2}\,,
\end{equation}
}where $g$ is the gravitational acceleration, and $ \omega_{\rm c}=c/2H$ is the acoustic cutoff frequency. The dispersion function (\ref{DR}) reduces to that set out in
equation (12) of \cite{sc06} in the 2D case $\phi=0$, with the addition of the decoupled
Alfv\'en factor $\omega^2-a^2\kpar^2$. It also takes the usual magneto\-acoustic
form $\omega^4-(a^2+c^2)\omega^2 k^2+a^2c^2k^2\kpar^2$ in the absence of
stratification, and the standard acoustic gravity wave form
$\omega^2(\omega^2- \omega_{\rm c}^2)+c^2N^2  k_{\rm h}^2$ without magnetic field. The first
term $\omega^2 \omega_{\rm c}^2 a_y^2  k_{\rm h}^2$ provides the coupling
between the magneto\-acoustic and Alfv\'en waves in three dimensions.

The dispersion relation $\calD=0$ restricts allowable solutions in $\k$-$\omega$ space. Now, in a horizontally and temporally invariant medium as is assumed here, $\omega$ and the horizontal component $k_x$ of $\k$ do not change with height, but the vertical component $k_z$ does. We may therefore fix $\omega$ and $k_x$ and plot dispersion curves in $z$-$k_z$ space. This gives a very informative overview of the propagation properties of the various modes, and in particular where they reflect.

\begin{figure}
\begin{center}
\includegraphics[width=0.799\hsize]{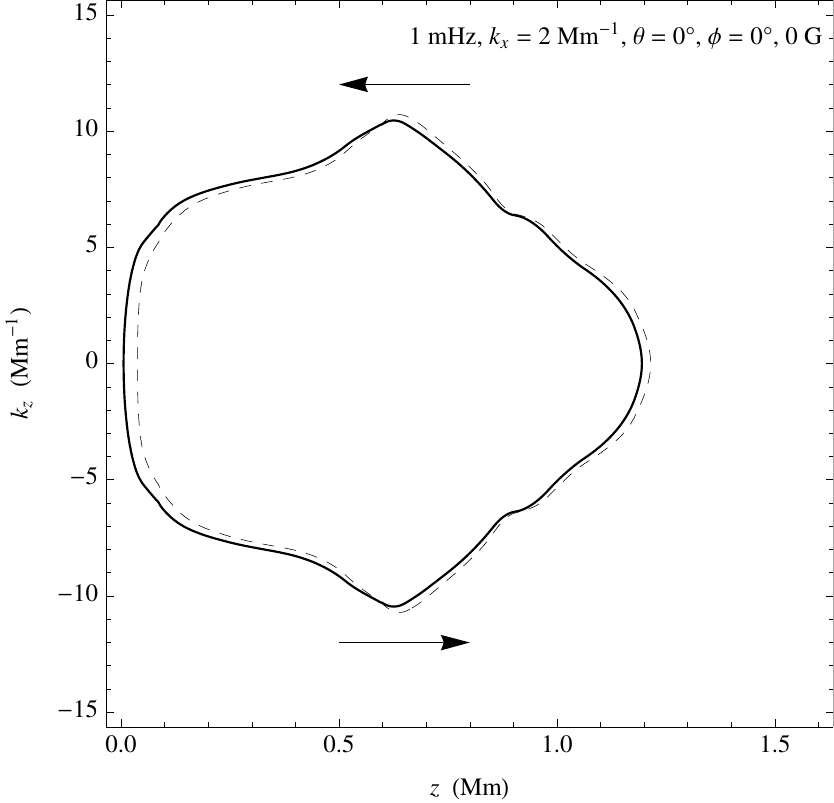}
\caption{$k_z$ vs.~$z$ dispersion diagram for the non-magnetic case at 1 mHz with $k_x=2$ $\rm Mm^{-1}$. The full curve represents $\calD=0$ with the isothermal acoustic cutoff formula, and the dashed curve is for the \citeauthor{dg84} formula. The lower branch represents a gravity wave with upward energy flux (as indicated by the arrow). Reflection occurs at $z\approx1.2$ Mm for this frequency and horizontal wavenumber.}
\label{fig:B0}
\end{center}
\end{figure}

Figure \ref{fig:B0} displays a representative propagation diagram for the VAL C atmosphere in the absence of magnetic field. As there is little difference in the curves with $\omega_{\rm ci}$ and $\omega_{\rm\scriptscriptstyle{DG}}$, the former will be used exclusively throughout. The dispersion relation reduces to
\begin{equation}
k_z^2 = \frac{N^2-\omega^2}{\omega^2}\,k_x^2 + \frac{\omega^2- \omega_{\rm c}^2}{c^2} \label{nonmag}
\end{equation}
in this case, clearly indicating vertical evanescence for $N<\omega< \omega_{\rm c}$ for all $k_x$. Even if $\omega<N< \omega_{\rm c}$, the wave is evanescent for small enough $k_x$. This explains why the gravity wave dispersion curve in Fig.~\ref{fig:B0} is trapped below about 1.2 Mm in a ``gravity cavity''. 

It is important to note that the upgoing branch in Fig.~\ref{fig:B0} is the \emph{lower} one, given the well-known property of gravity waves that the vertical components of the phase and group velocities have opposite signs {\hilite\citep{light78}}. Thus gravity waves with positive upward energy flux have downward phase velocity, \emph{i.e.}, $k_z<0$.

%%%%%
\subsubsection{Low $\beta$ Asymptotics}
The asymptotic behaviour of wave modes in the $a\gg c$ region is of crucial importance. {\hilite A wide description of the behaviour of slow waves in this regime is given by \cite{roberts06}, including a discussion of applications to coronal loops. Magnetically structured media (\emph{e.g.}, flux tubes) are beyond the scope of this paper, and will not be considered here. }

Analysis of the dispersion relation in the low-$\beta$ regime reveals that there are two (possible) oscillatory classes of solution:
\begin{itemize}
\item Field-aligned acoustic (slow) waves $\omega^2\sim c^2\kpar^2+ \omega_{\rm c}^2\cos^2\theta$, \emph{i.e.},
\begin{equation}
k_z \sim \pm \frac{\sqrt{\omega^2- \omega_{\rm c}^2\cos^2\theta}}{c\,\cos\theta} -\tan\theta\cos\phi\, k_x \mbox{\ \ as $a\to\infty$,}\label{ac_kz}
\end{equation}
provided $\omega> \omega_{\rm c}\cos\theta$. We term the reduction of the effective acoustic cutoff frequency by the factor $\cos\theta$ the ``ramp effect''. Equation (\ref{ac_kz}) agrees perfectly with the exact $\mu_3$ Frobenius eigenvalue in equation (26) of \cite{cg08}.
\item The Alfv\'en wave (which is also field-aligned of course), 
\[
\omega^2\sim a^2\kpar^2+
\frac{\omega^2 \omega_{\rm c}^2}{\omega^2- \omega_{\rm c}^2\cos^2\theta}\,  \frac{\sin^2\theta\sin^2\phi}{1+\tan^2\theta\cos^2\phi}
\]
 \emph{i.e.},
\begin{multline}
k_z \sim \pm \frac{\omega\sec\theta}{a}\sqrt{1-
\frac{ \omega_{\rm c}^2}{\omega^2- \omega_{\rm c}^2\cos^2\theta}\,  \frac{\sin^2\theta\sin^2\phi}{1+\tan^2\theta\cos^2\phi}}  \\[4pt]
{}-\,\tan\theta\cos\phi\, k_x \, ,\label{alf_kz}
\end{multline}
also as $a\to\infty$.
\end{itemize}
In both cases, the term $k_z=-\tan\theta\cos\phi\, k_x$ alone would keep the phase of $\exp[\irom(k_xx+k_zz)]$ constant on field lines $x=z\tan\theta\cos\phi+$constant, so it is the remaining square root terms which determine phase along field lines for these two field-aligned wave modes. {\hilite In other words, $-\tan\theta\cos\phi\, k_x$ disappears if $z$ is redefined as height along a fixed field line rather than at fixed $x$. Such a term is expected in the low-$\beta$ regime where the rigidity of the magnetic field lines dominates. It is purely geometric.}

The remaining fast magneto-acoustic wave is always evanescent {\hilite for fixed $\omega$ and $k_x$ and large enough $a\gg c$ since its dispersion relation is $k_z^2\approx \omega^2/a^2-k_x^2<0$ }.

%The Alfv\'en result (\ref{alf_kz}) should not be taken too literally though. The eikonal approximation at the heart of the derivation of the dispersion relation assumes that $k_zH\gg 1$ formally, which of course breaks down as $a$ increases. The dispersion relation's description of Alfv\'en waves is therefore indicative rather than definitive at low $\beta$ (though we shall see that, as is common, the dispersion relation often performs better than it has any right to). We shall rely on detailed numerical integration to quantitatively calculate the Alfv\'en wave energy flux in the low $\beta$ region.

%%%%%%%%%%%%%%
\subsection{Numerical Integration}   \label{sub:num}
Although convenient and informative, the dispersion relation is only an approximate tool. We corroborate our results by numerically integrating the linear adiabatic wave equations across $0<z<1.6$ Mm. The integrated equations do not depend upon $ \omega_{\rm c}$ explicitly, and so are not subject to the uncertainties associated with that term. The top boundary conditions are taken from the exact convergent series solutions derived by \cite{cg08} {\hilite for isothermal slabs}, where $a\gg c$, and which represent (i) an outgoing or evanescent slow (acoustic) wave (depending on whether $\omega> \omega_{\rm c}\cos\theta$ or \emph{vice versa}; (ii) an evanescent fast (magnetic) wave; and (iii) an outgoing Alfv\'en wave (in the 3D case). There are no incoming or exponentially growing solutions allowed at the top.

At the bottom ($z\leq0$) we generalize the radiation boundary condition (8) of \cite{cg08} to suit gravity waves as well as acoustic waves. The aim is to disallow incoming slow (magnetic) and Alfv\'en waves at the base, where $a\ll c$, but allow incoming gravity or acoustic waves, and outgoing waves of all varieties. We do this by imposing
\begin{equation}
\left(\pderiv{}{t}-\a\vdot\grad\right)\mathcal{A}\,\bxi=0
\quad\mbox{and}\quad
\left(\pderiv{}{t}-\a\vdot\grad\right)\rho^{1/4}\eta=0
\, ,  \label{radn}
\end{equation}
where $\bxi=(\xi,\eta,\zeta)$ is the plasma displacement vector (though the $\eta$ equation is redundant in 2D where $\eta$ {\hilite decouples}), and $\mathcal{A}$ is the gravito-acoustic annihilator operator defined by
\begin{equation}
\mathcal{A}\,\bxi = \rho^{-1/4}\left[\pderiv{\xi}{z}-\irom\,k_x\zeta -\frac{N^2\xi}{g}+\irom\, k_x \frac{N^2\zeta}{\omega^2}\right]\, .  \label{annih}
\end{equation}
The square bracket term in equation (\ref{annih}) is identically zero in the absence of magnetic field. Therefore, in the region $a\ll c$ where the gravito-acoustic and magnetic waves are fully decoupled, it {\hilite strongly suppresses} the gravito-acoustic waves, leaving only the slow (magnetic) waves.  The factor $\rho^{-1/4}$ in (\ref{annih}) removes an amplitude factor in the remaining sinusoidal slow waves. The $\partial/\partial t - \a\vdot\grad$ operator then selects the upcoming magnetic wave, which is set to zero with this condition. The incoming Alfv\'en wave is similarly suppressed by the second equation in (\ref{radn}).\footnote{\hilite The $\rho^{\pm 1/4}$ factors may be unfamiliar, and so warrant some explanation. The Alfv\'en wave equation
takes the form $\partial^2\eta/\partial t^2=a^2 \partial^2\eta/\partial s^2$, where $s=z\sec\theta$ is distance along the field line and $a\propto\rho^{-1/2}$ is the Alfv\'en speed. Assuming harmonic time dependence $\exp(-\irom\omega t)$ reduces this to $\partial^2\eta/\partial z^2=Q(z)\eta$, where $Q=-\omega^2/(a^2\cos^2\theta)\propto\rho$. The WKBJ amplitude factor for such an equation is then $|Q|^{-1/4}\propto\rho^{-1/4}$ \citep{bo}. Thus the $\rho^{1/4}$ factor in (\ref{radn}) expunges this amplitude dependence to leading order. The slow wave in $a\ll c$ takes essentially the same form, though the polarization is in the $x$ rather than $y$ direction, and $\xi$ may be used as the dependent variable instead of $\eta$. The dominant term in the square bracket in $\mathcal{A}\bxi$ is obviously $\partial\xi/\partial z\propto k_z\xi$. But $k_z\sim \omega/a\cos\theta\propto\rho^{1/2}$ and the amplitude of $\xi$ varies as $\rho^{-1/4}$ as before, showing that $\partial \xi/\partial z$ has amplitude factor $\rho^{1/4}$ to leading order. The $\rho^{-1/4}$ factor in (\ref{annih}) cancels this. These arguments are confirmed numerically.}

\begin{figure}
\begin{center}
\includegraphics[width=\hsize]{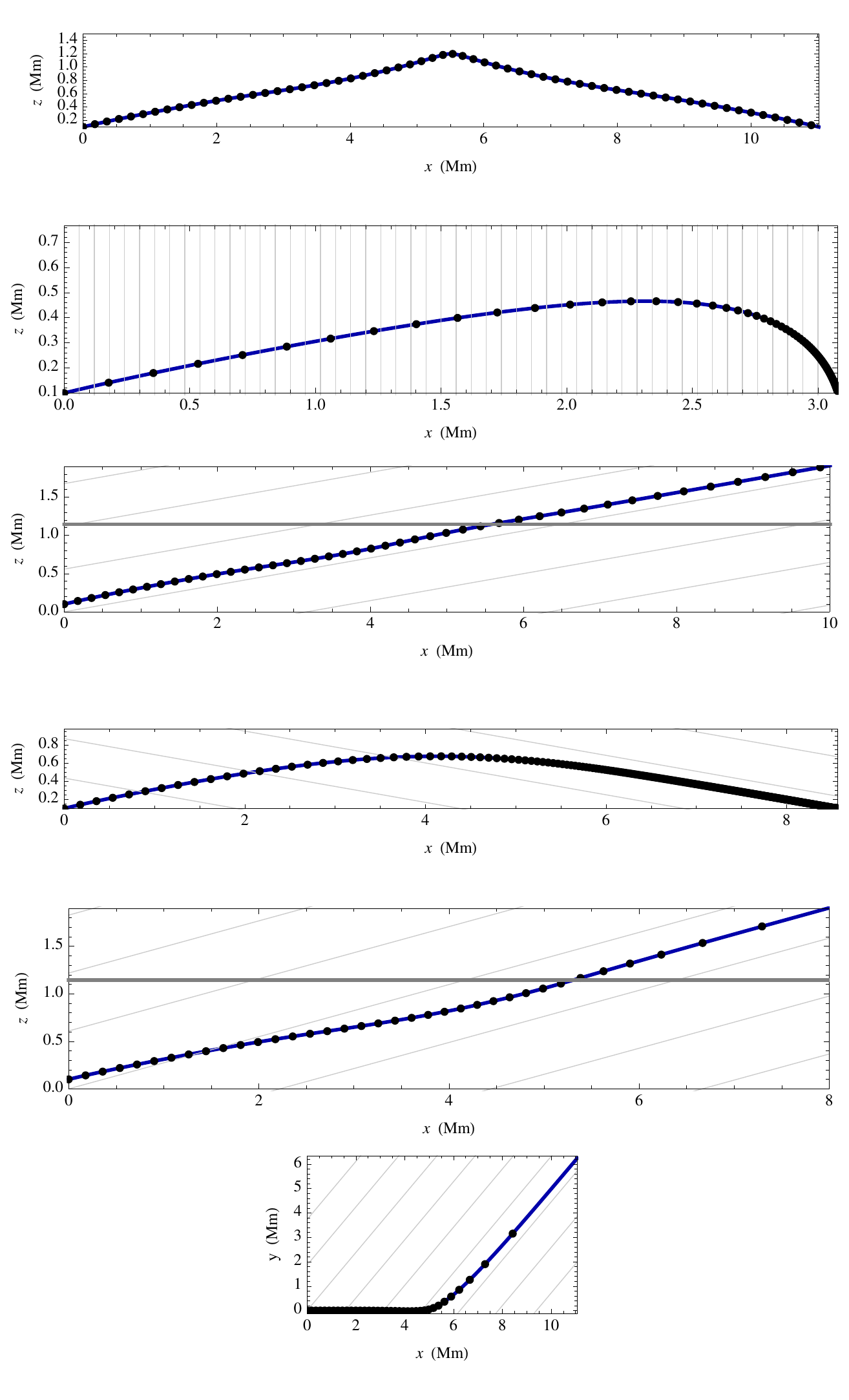}
\caption{Ray plots for rays of frequency 1 mHz and horizontal wavenumber $k_x=2$ $\rm Mm^{-1}$ with (from top to bottom): (i) no magnetic field; (ii) vertical 10 G field; (iii) 10 G field inclined $80^\circ$ from the vertical (2D); (iv) 10 G field inclined $-80^\circ$ from the vertical; (v) 10 G field inclined $80^\circ$ from the vertical and oriented $50^\circ$ away from the $x$-$z$ plane (3D); and (vi) $x$-$y$ projection of the previous case. Field line orientation is indicated by the faint grey backgrounds (projected in case (v)). The dots on the ray curves indicate 1 minute group travel intervals. The $a=c$ equipartition level is indicated by a horizontal line where it falls within the depicted domain. In all cases the rays are propagating {\hilite rightward}.}
\label{fig:rays}
\end{center}
\end{figure}

The differential equations being of $4^{\rm th}$ ($6^{\rm th}$) order in 2D (3D), and with 2 (3) boundary conditions applied at the top and 1 (2) at the bottom, an arbitrary normalization condition is all that remains to be applied. The gravity waves are assumed generated at $z=0$ (by granulation), so it is not necessary to impose an evanescence condition on them below this level, which would result in an eigenvalue problem, rather than the driven wave scenario we envisage here.

As a test of the bottom boundary condition, we apply it at several depths ($z=0,\,-1,\,-2$ Mm), and find that our solution typically varies only in the third significant figure. This indicates that equations (\ref{radn}) represent a good approximation to the required radiation condition. One could do better by developing an asymptotic series solution \citep[as in][]{CC05} and applying it at sufficient depth, but this is both messy and unnecessary, it would be specific to the particular model atmosphere in which it were applied, and it could not {\hilite easily} be done for a tabulated model.

%%%%%%%%%%%%%%%%%%%%%%%%%%%%%%%%%%%%%%
\section{RESULTS}   \label{results}

For simplicity, we focus on 1 mHz waves with horizontal wavenumber $k_x=2$ $\rm Mm^{-1}$, placing them well within the high flux regime identified observationally by \cite{straus08}. Two magnetic field strengths, $B=10$ G and $B=100$ G will suffice to illustrate how the gravity waves are influenced by typical chromospheric fields.

%%%%%%%%%%%
\subsection{Dispersion and Ray Diagrams}

\begin{figure*}
\begin{center}
\includegraphics[width=0.799\hsize]{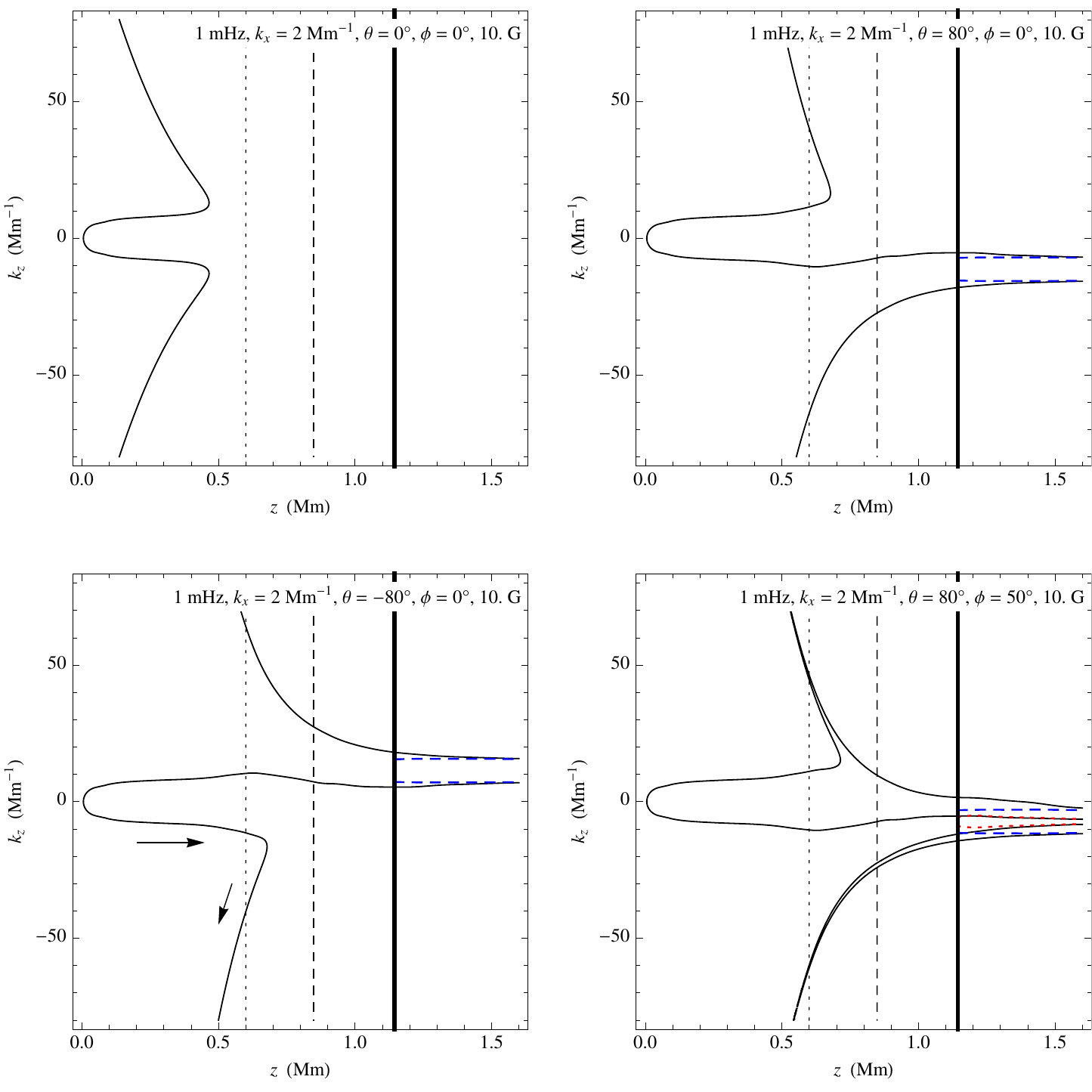}
\caption{Dispersion diagrams for waves of frequency 1 mHz and horizontal wavenumber $k_{x}$=2 Mm$^{-1}$. The vertical lines indicate the value of the ratio $a^{2}/c^{2}$ at various heights: thick solid line -- $a^{2}/c^{2}=1$ (the equipartition level); dashed line -- $a^{2}/c^{2}=0.1$;  dotted line --  $a^{2}/c^{2}=0.01$. Top left:  vertical 10 G field -- the presence of even a weak vertical magnetic field causes the up-going gravity wave (bottom branch) to reflect back downwards as a slow magneto-acoustic wave, and it does so even before $a^{2}/c^{2}$ reaches 1\% of the equipartition value; Top right: 10 G field inclined $80^\circ$ from the vertical (2D) -- when the field is highly inclined and the attack angle is small, the up-going gravity wave penetrates the equipartition level and connects to the up-going field-aligned acoustic wave (given asymptotically by equation (\ref{ac_kz}) which is indicated here by the blue dashed lines); Bottom left: 10 G field inclined $-80^\circ$ from the vertical -- when the attack angle is large, the up-going gravity wave is reflected down as a slow magneto-acoustic wave (as indicated by the arrows); Bottom right 10 G field inclined $80^\circ$ from the vertical and oriented $50^\circ$ away from the $x$-$z$ plane (3D) -- in this case the up-going gravity wave connects to the Alfv\'en  solution (asymptotic equation (\ref{alf_kz}), the red dotted curve). These results are in complete agreement with the ray behaviours of Fig.~\ref{fig:rays}. }
\label{fig:dispdgrid}
\end{center}
\end{figure*}

Ray plots\footnote{The ray equations are derived from the dispersion function, which takes the role of a Hamiltonian. See for example \cite{cally09a}, equation (14). They are solved as a set of ordinary differential equations along their arc lengths, starting at $x=0$, $z=0.1$, just inside the gravity cavity.} for four cases are collected in Fig.~\ref{fig:rays}. They illustrate, respectively,
\begin{enumerate}
\item In the absence of magnetic field, the gravity wave reflects symmetrically at around $z=1.2$ Mm, as expected from Fig.~\ref{fig:B0}.
\item In vertical field, the wave reflects at a much lower height, and as a slow magneto-acoustic wave (we can see that it is ``slow'' by the very close spacing of the 1 min dots).
\item At high field inclination in 2D, the gravity wave converts to an acoustic wave which then follows the field lines. The regular grid spacing above $a=c$ indicates that it is acoustic (the atmosphere is nearly isothermal).
\item At high inclination in the other direction (negative $\theta$) the ray again reflects as a slow wave following the field lines.
\item At high field inclination in 3D, with $\phi=50^\circ$ here, the mode conversion is to the Alfv\'en wave, as is apparent from the rapidly increasing 1-min dot separation with height. The depicted ray path is a projection onto the $x$-$z$ plane. In reality, this last ray turns sharply out of the $x$-$z$ plane to follow the field lines, as illustrated in the bottom panel.
\end{enumerate}
Clearly, the determining characteristic here for the different behaviours is the ``attack angle'' \citep{sc06} the ray makes to the magnetic field lines. At small attack angle the ray may avoid reflection.
 
These are ``classical'' ray calculations \citep{wein}, as distinct from the generalized ray formalism of \cite{sc06} \citep[recently tested against exact solutions by][]{hc09}. Therefore, they do not account for partial transmission/conversion. Instead the rays simply follow the connectivity implied by the dispersion curves. Nevertheless, these are often (but not always) the dominant paths. Partial mode conversions are addressed in Section \ref{subsec:flux} using full numerical solution.

Fig.~\ref{fig:dispdgrid} displays the dispersion diagrams corresponding to the ray paths depicted in Fig.~\ref{fig:rays}, parts (ii)--(v) respectively. The non-magnetic case (part (i)) has already been covered in Fig.~\ref{fig:B0}. Points to note include (with numbering corresponding to that of the previous list)
\renewcommand{\labelenumi}{(\Roman{enumi})}
\begin{enumerate}
\setcounter{enumi}{1}
\item The dispersion curve turns over at a very low height in vertical magnetic field, even before $a^2/c^2$ reaches the 1\% level, in complete agreement with the corresponding ray path. The ``return path'' clearly corresponds to very slow propagation, as $|k_z|$ becomes very large (recall that the phase speed is $\omega/|\k|$ and $\omega$ and $k_x$ are fixed).
\item With highly inclined 2D field the upgoing gravity branch penetrates the $a=c$ equipartition level and quickly settles onto the low-$\beta$ asymptotic acoustic curve (blue dashed lines). Again, this is in total accord with the ray figure.
\item When the highly inclined ($80^\circ$) field is oppositely oriented ($-80^\circ$) and the attack angle is large, there is no such connection to the low-$\beta$ acoustic waves, and the gravity wave reflects once again as a slow magneto-acoustic wave, as indicated by the arrows in the lower left panel of Fig.~\ref{fig:dispdgrid}. Note that this panel is simply the reflection about $k_z=0$ of the top right panel.
\item In the 3D case the Alfv\'en wave now couples to the acoustic-gravity waves (the Alfv\'en loci have been suppressed in the first three panels, as they were inaccessible in 2D). In the case depicted, the upgoing gravity wave connects directly to the Alfv\'en wave, as indicated by the red dotted curve representing the Alfv\'en asymptotics of equation (\ref{alf_kz}). Once again, the agreement with the ray figure is apparent.
\end{enumerate}

%%%%%%%%%%%
\subsection{Transmitted Fluxes}   \label{subsec:flux}

\begin{figure}
\begin{center}
\includegraphics[width=0.799\hsize]{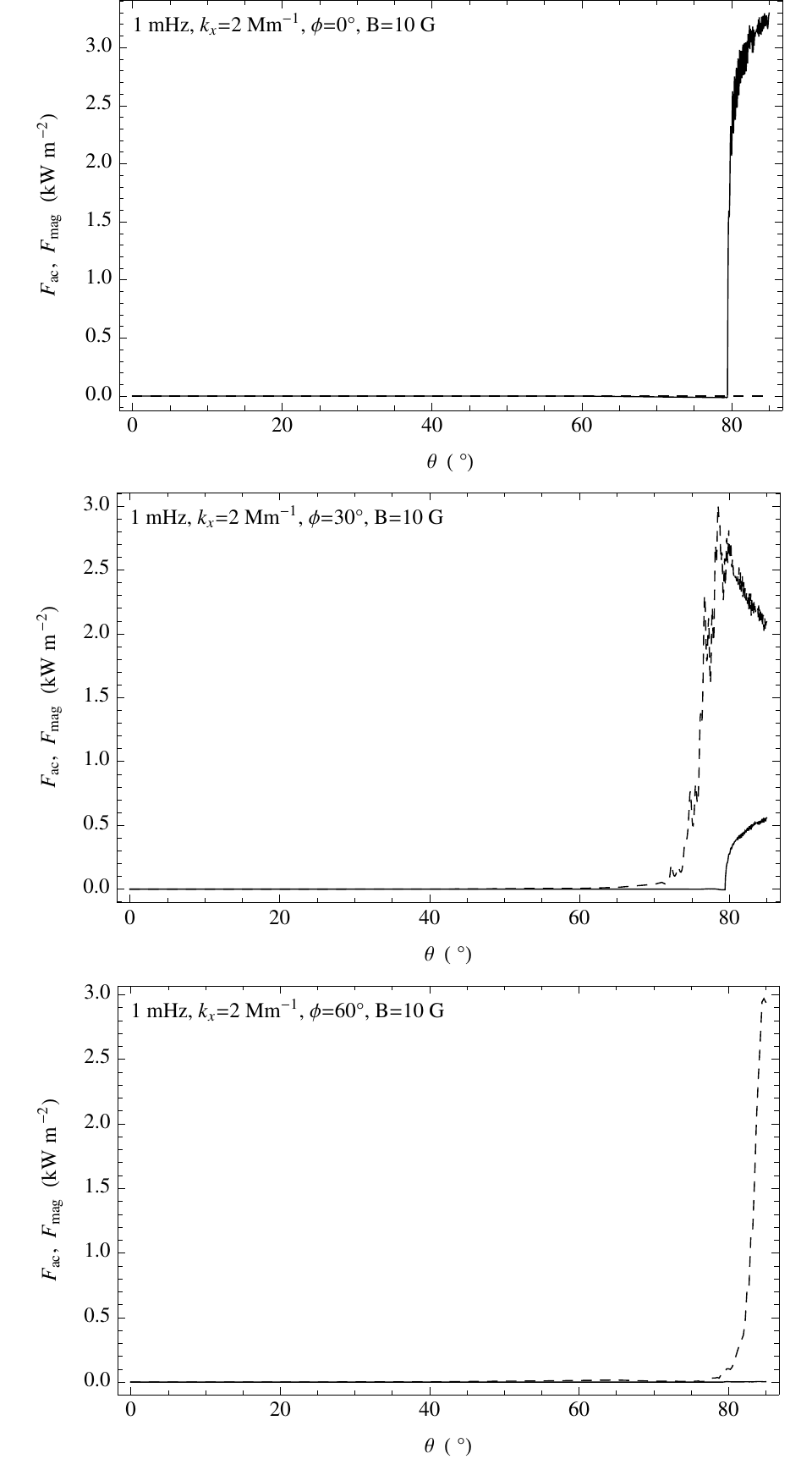}
\caption{Acoustic (full) and magnetic (dashed) wave energy flux ($\rm kW\,m^{-2}$) as functions of magnetic field inclination $\theta$ and three orientations $\phi$ {\hilite ($0^\circ$, $30^\circ$, and $60^\circ$ respectively)} for $B=10$ G and 1 mHz waves with $k_x=2$ $\rm Mm^{-1}$. In all cases, the vertical velocity $w$ at $z=0$ is normalized to 1 $\rm km\,s^{-1}$. Despite appearances, the rapid oscillations at high inclination are smoothly resolved, and are not numerical artefacts.}
\label{fig:F10theta}
\end{center}
\end{figure}

\begin{figure}
\begin{center}
\includegraphics[width=0.799\hsize]{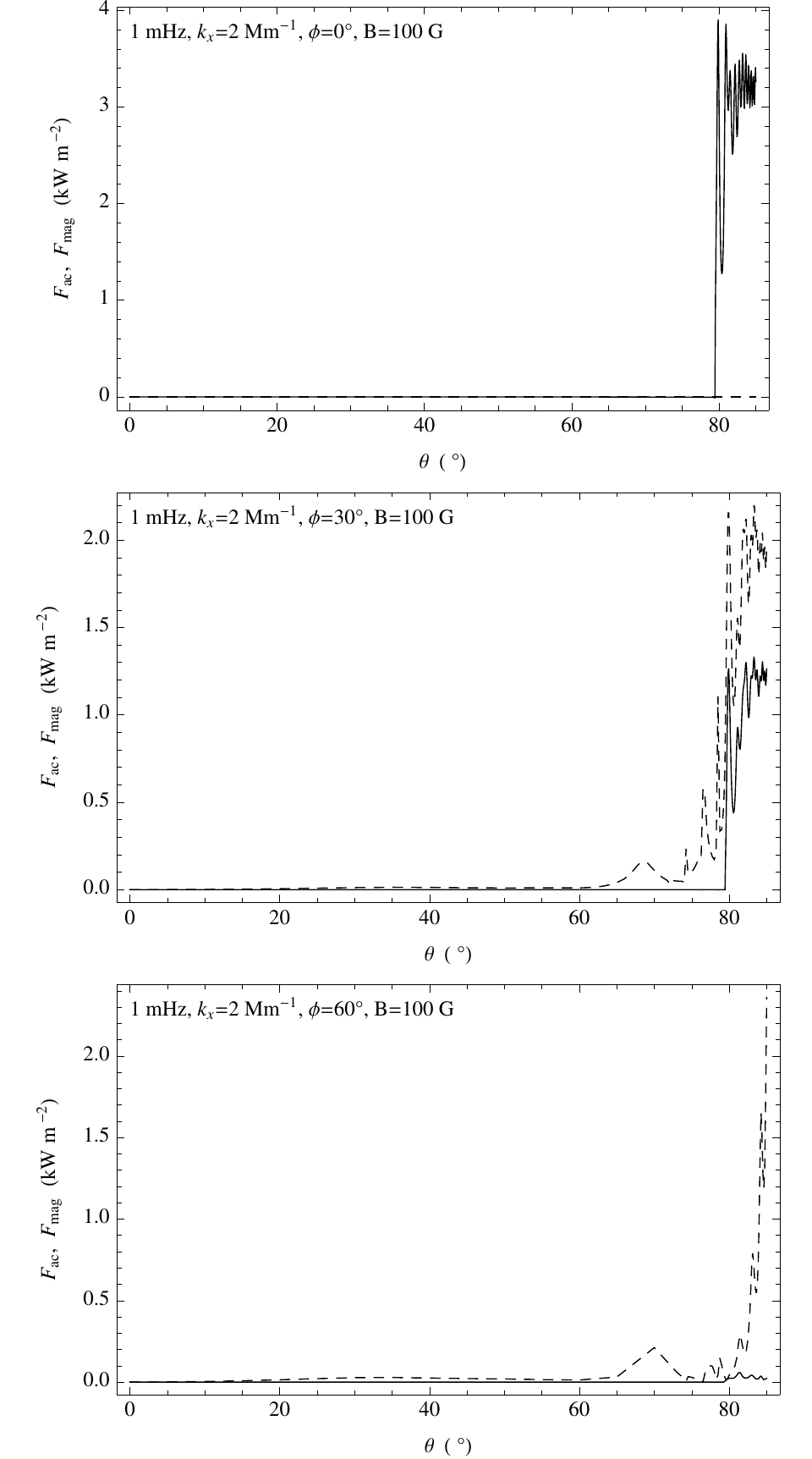}
\caption{As for Fig.~\ref{fig:F10theta}, but with $B=100$ G.}
\label{fig:F100theta}
\end{center}
\end{figure}

\begin{figure}
\begin{center}
\includegraphics[width=0.799\hsize]{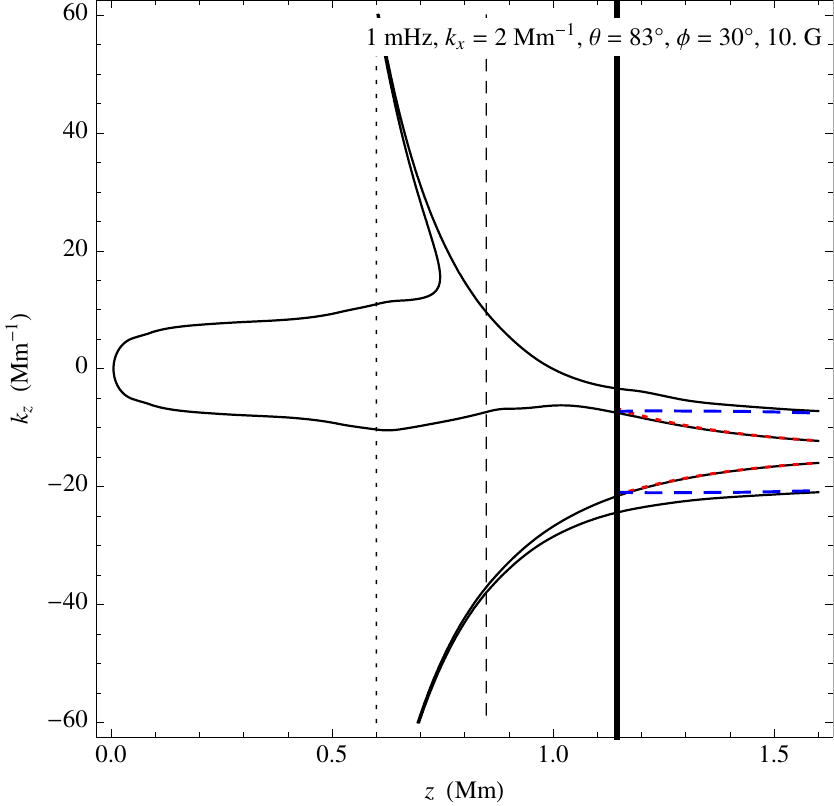}
\caption{Dispersion diagram for a wave with frequency 1 mHz and horizontal wavenumber $k_{x}$=2 Mm$^{-1}$, subject to a 10 G field with $\theta=83^{\circ}$, $\phi=30^{\circ}$. The gravity wave dispersion curve connects to the Alfv\'en wave above $a=c$. However the field guided acoustic wave (asymptotic to the upper blue dashed line) passes close to this gravity/Alfv\'en locus near the equipartition level, and energy can tunnel across the gap (\emph{c.f.} the middle panel of Fig.~\ref{fig:F10theta}).}
\label{fig:3Dtunnelling}
\end{center}
\end{figure}

{\hilite 
The wave-energy flux carried by linear MHD waves may be broken into gravito-acoustic and magnetic parts, $\F=\F_{\rm ac}+\F_{\rm mag}$, where $\F_{\rm ac}=\Re[p_1\v^\ast]$ and $\F_{\rm mag}=-\Re[\E_1\vcross\B_1^\ast]$, $p_1$ is the Eulerian gas pressure perturbation, $\v$ is the plasma velocity, and $\E_1$ and $\B_1$ are the electric and magnetic field perturbations. Once the numerical solution is obtained, the vertical components $F_{\rm ac}$ and $F_{\rm mag}$ may be calculated in the isothermal layer at the top \citep[see][for details]{cally09b}.} In 3D, the flux of most interest is $F_{\rm mag}$, the Alfv\'en flux escaping from the top (identified as such since the fast wave is evanescent in that regime). In all cases, the vertical velocity $w$ at $z=0$ is normalized to 1 $\rm km\,s^{-1}$. Since the flux scales as the square of the velocity perturbations, these results are easily adjusted for different driving velocities.

Fig.~\ref{fig:F10theta} for $B=10$ G shows substantial acoustic flux in the 2D case $\phi=0^\circ$, once the ramp effect has {\hilite turned on} ($\cos\theta<\omega/ \omega_{\rm c}$). In practice, this requires highly inclined field, characteristic of canopy. As the field is rotated out of the $x$-$z$ plane, coupling to the Alfv\'en wave takes over, and the acoustic flux diminishes. However, again it is only significant at high field inclination. Very similar results are found at 100 G (Fig.~\ref{fig:F100theta}). It is most interesting that the resultant fluxes, even with a modest base driving velocity of $w=1$ $\rm km\,s^{-1}$, are comparable to the estimated chromospheric radiative losses, and are therefore of genuine significance.

It is important to not give too much weight to the wave connectivities implied by the eikonal approximation through the dispersion and ray diagrams. As is familiar from quantum mechanics, energy can jump gaps between (closely) neighbouring dispersion curves. This is well illustrated by Fig.~\ref{fig:3Dtunnelling}. The second panel of Fig.~\ref{fig:F10theta} shows that both acoustic and Alfv\'enic flux exits the top in this 3D case, though with the magnetic flux dominating. Comparing with the dispersion curve, the eikonal connectivity is to the Alfv\'en branch, but the gap (avoided crossing) to the acoustic branch is narrow. This is clearly indicative of partial tunnelling across the gap.

\begin{figure*}
\begin{center}
\includegraphics[width=0.799\hsize]{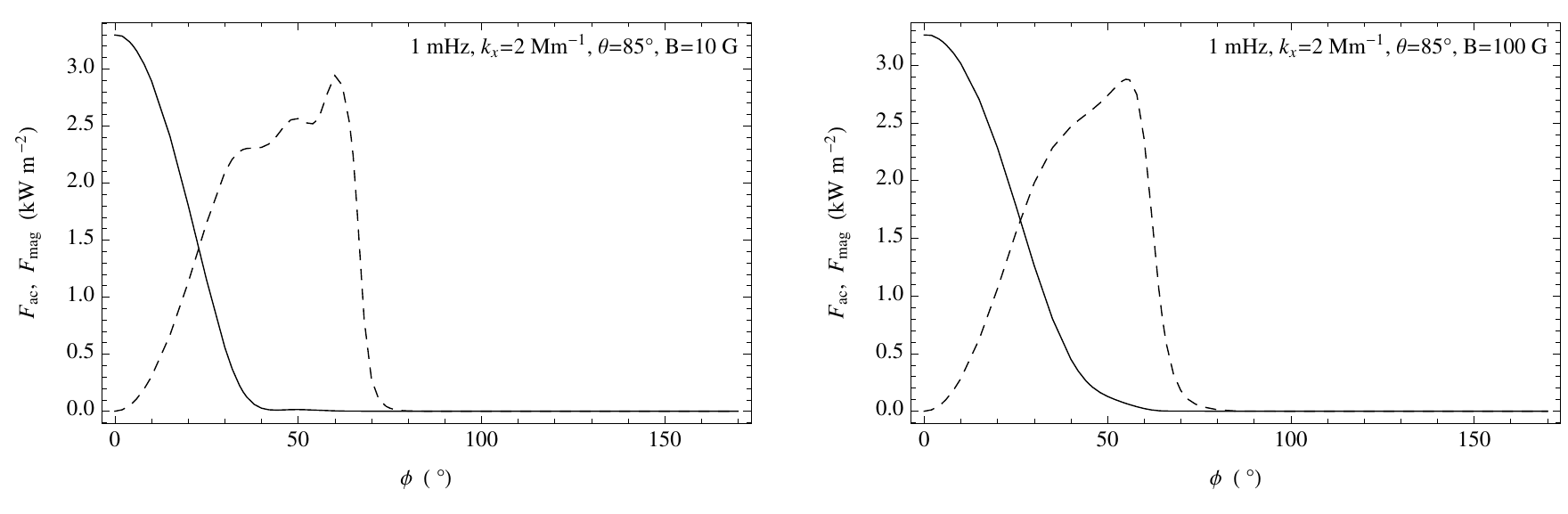}
\caption{Acoustic (full) and magnetic (dashed) wave energy flux ($\rm kW\,m^{-2}$) as functions of magnetic field orientation $\phi$ with inclination $\theta=85^\circ$. Left panel: $B=10$ G; right panel: $B=100$ G. As in figures \ref{fig:F10theta} and \ref{fig:F100theta}, the frequency is 1 mHz and the horizontal wavenumber is $k_x=2$ $\rm Mm^{-1}$. The vertical velocity $w$ at $z=0$ is normalized to 1 $\rm km\,s^{-1}$.}
\label{fig:Fphi}
\end{center}
\end{figure*}

\begin{figure*}
\begin{center}
\includegraphics[width=\hsize]{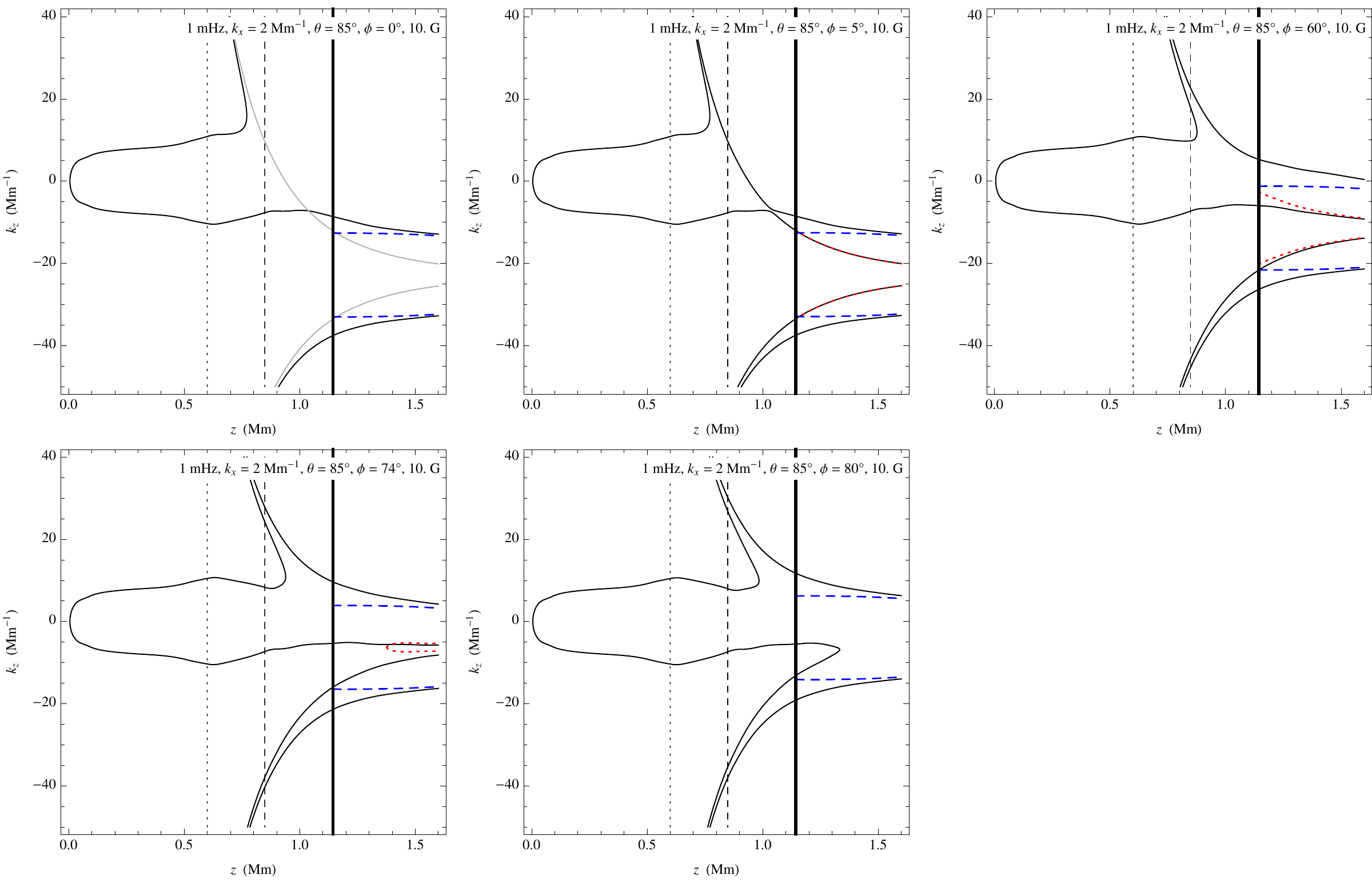}
\caption{{\hilite Dispersion diagrams illustrating the role of field orientation $\phi$ in the waxing and waning of Alfv\'en energy loss. All cases correspond to field inclination $\theta=85^\circ$, field strength 10 G, frequency 1 mHz, and horizontal wavenumber 2 $\rm Mm^{-1}$. Top left: $\phi=0^\circ$, with the Alfv\'en branch greyed out to denote that it is decoupled from the gravito-magneto-acoustic branches; Top centre: $\phi=5^\circ$; Top right: $\phi=60^\circ$; Bottom left: $\phi=74^\circ$; and Bottom centre: $\phi=80^\circ$.}}
\label{fig:Alf}
\end{center}
\end{figure*}

Fig.~\ref{fig:Fphi} fixes $\theta=85^\circ$ and rotates $\phi$ from $0^\circ$ to $170^\circ$. Once again, we see that the acoustic flux diminishes as the field orientation moves away from the 2D configuration, but that it is largely compensated by an increase in Alfv\'en flux, at least out to about $70^\circ$, after which both fluxes quickly become insignificant. This is reminiscent of Fig.~2 in \cite{cg08}, though shifted here to much higher $\theta$ because of the lower frequency. {\hilite  Fig.~\ref{fig:Alf} illustrates this further using dispersion diagrams. In 2D the Alfv\'en branch is inaccessible, and so carries no flux. However, as the magnetic field is rotated out of the $x$-$z$ plane, the upgoing gravity wave branch instantly connects to the Alfv\'en branch at large $z$. Most energy continues to flow along the low-$\beta$ acoustic branch though by jumping the (small) gap that has opened up as an avoided crossing (primarily because the polarizations are still almost orthogonal at small $\phi$). At larger $\phi$ ($60^\circ$ and $74^\circ$ are illustrated here) the gap is very large and so the energy flows overwhelmingly along the dispersion curve onto the Alfv\'en branch. However, by $\phi=80^\circ$ this upgoing Alfv\'en branch has been cut off (due to the term under the square root in (\ref{alf_kz}) becoming negative) and the connectivity is now to a downgoing Alfv\'en wave, explaining the precipitous drop in Alfv\'en flux with increasing $\phi$ seen in Fig.~\ref{fig:Fphi}. }

\begin{figure}
\begin{center}
\includegraphics[width=0.799\hsize]{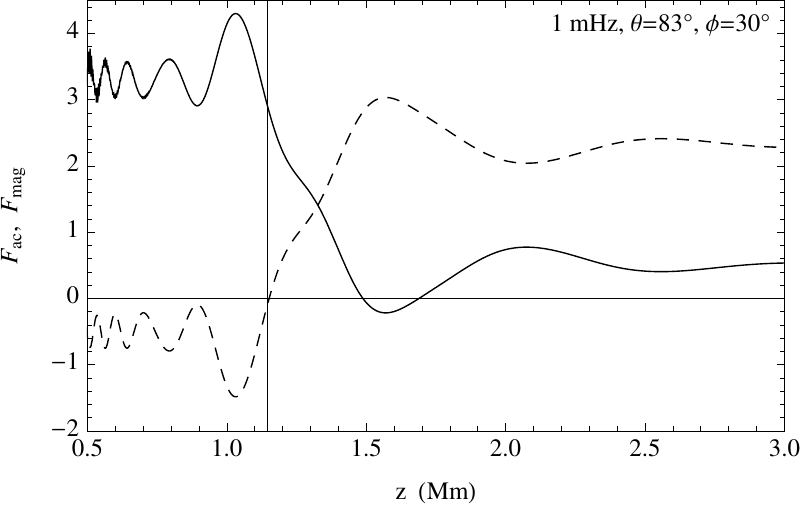}
\caption{Acoustic (full) and magnetic (dashed) wave energy fluxes as functions of height $z$ (Mm) for $B=10$ G, $\theta=83^\circ$, $\phi=30^\circ$, corresponding to Fig.~\ref{fig:3Dtunnelling}. Once again, the frequency is 1 mHz and the horizontal wavenumber is 2 $\rm Mm^{-1}$. The vertical line indicates the position of the Alfv\'en/acoustic equipartition level $a=c$. An extended section of the isothermal}
\label{fig:Fz}
\end{center}
\end{figure}

Fig.~\ref{fig:Fz} illustrates how wave energy flux may shift from acoustic to magnetic near the $a=c$ equipartition level in 3D. Also note the \emph{negative} magnetic flux at the bottom, indicating some significant reflected Alfv\'en and/or slow magneto-acoustic flux there. Typically, the conversion to Alfv\'en waves occurs over a broader and slightly higher region than is characteristic of fast/slow conversion, and that is apparent here, with a steady conversion process taking place between roughly 1 Mm and 1.5 Mm. The fact that the dispersion locus which penetrates $a=c$ in Fig.~\ref{fig:3Dtunnelling} essentially sits right on top of the asymptotic Alfv\'en curve (red dotted) shows just how well the acoustic-gravity and Alfv\'en phase velocities match over an extended interval, which is very favourable for mode conversion \citep{cally05}.

Fig.~\ref{fig:Ffreqs} briefly addresses how the escaping wave energy fluxes at the top vary with frequency. The two cases shown are for 0.7 mHz and 2.1 mHz, but otherwise address the same case as in the middle panel of Fig.~\ref{fig:F10theta}. The results are qualitatively similar, except that the flux ``turns on'' at greater inclinations for lower frequency, and lesser inclinations for higher frequency, as expected from the ramp effect, which allows acoustic waves to propagate once $\omega> \omega_{\rm c}\cos\theta$. It is also notable that the magnitude of the Alfv\'enic flux is considerably increased at higher frequency.

%%%%%%%%%%%%%%%%%%%%%%%%%%%%%%%%%%%%%%%
\section{CONCLUSIONS}     \label{conclusions}
\begin{figure}
\begin{center}
\includegraphics[width=0.799\hsize]{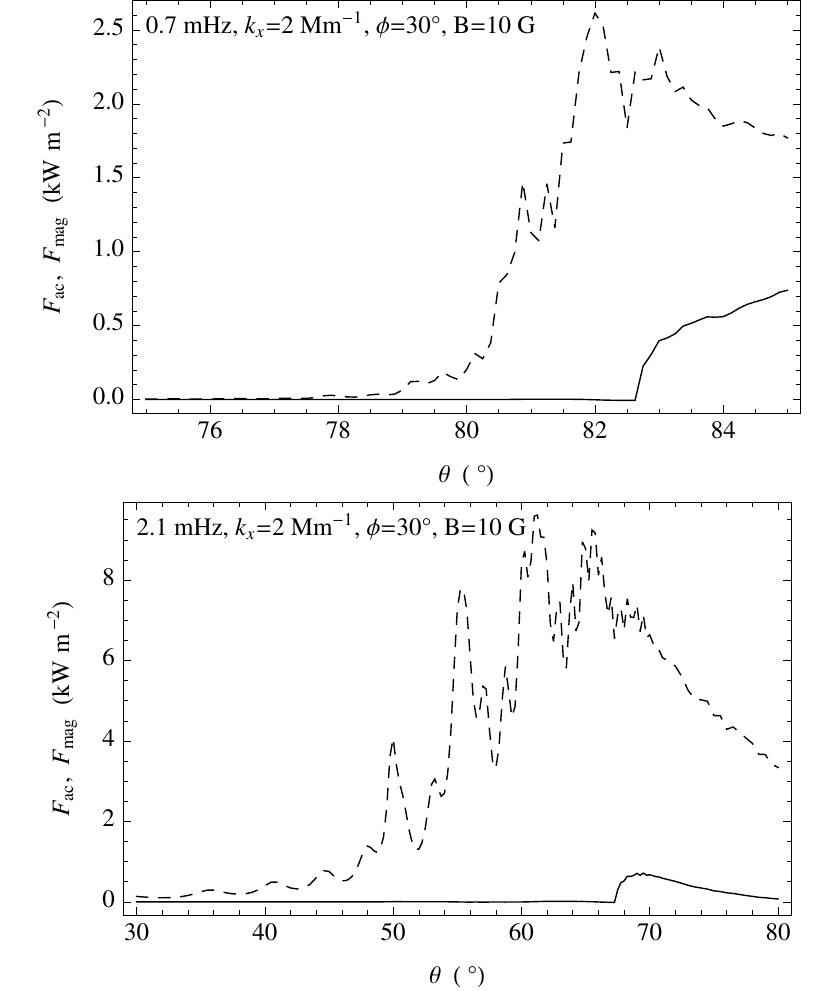}
\caption{Acoustic (full) and magnetic (dashed) wave energy flux ($\rm kW\,m^{-2}$) as functions of magnetic field inclination $\theta$ with $\phi=30^\circ$ for $B=10$ G, $k_x=0.2$ $\rm Mm^{-1}$, and two different frequencies. Top panel: 0.7 mHz; Bottom panel: 2.1 mHz. The middle panel of Fig.~\ref{fig:F10theta} is intermediate between these two cases. Note the different $\theta$-scales on the two graphs.}
\label{fig:Ffreqs}
\end{center}
\end{figure}

The main conclusions we draw from our analyses are:
\renewcommand{\labelenumi}{\arabic{enumi}.}
\begin{enumerate}
\item Even very weak magnetic fields very effectively reflect gravity waves back downward as slow magneto-acoustic waves. This typically happens well below the $a=c$ equipartition level.
\item However, at very large magnetic field inclinations, typically around $80^\circ$ or more depending on frequency (see Fig.~\ref{fig:Ffreqs}), substantial mode conversion from gravity waves to either field-guided acoustic waves (for small $\phi$) or Alfv\'en waves ($20^\circ\la \phi \la 70^\circ$) occurs, and these waves continue to propagate upward along the field lines. The amount of energy they carry is potentially significant for the upper chromosphere.
\item Wave energy fluxes reaching the top of our model are very sensitive to magnetic field direction, but quite insensitive to magnetic field strength, at least in the range 10 G -- 100 G.
\item The dispersion diagrams give a simple, easy and quite accurate picture of the behaviour of gravity waves in a magneto-atmosphere, though with the caveat that tunnelling can sometimes occur between branches.
\end{enumerate}

In simple terms, we conclude that atmospheric gravity waves are very effectively suppressed by even very weak magnetic field, \emph{unless} that field is highly inclined and the attack angle fine, in which case it opens a window to the upper atmosphere that allows the gravity waves to propagate through in a different guise. This is closely related to the ``magnetic portals'' of \cite{jeff06} which rely on the ramp effect to allow acoustic waves below the acoustic cutoff frequency to still propagate upward in low-$\beta$ inclined magnetic field, but in the case of low-frequency gravity waves which are already propagating, it gives them the opportunity to convert to propagating acoustic waves around the $a=c$ level. Because of their low frequency though, very substantial inclination is required to open these windows, but no more than is characteristic of chromospheric canopy.

{\hilite 
There are consequences for recent and future observations of solar atmospheric oscillations. The surprising extent to which even very weak vertical or moderately inclined magnetic field inhibits gravity waves by causing them to quickly reflect as slow magneto-acoustic waves perhaps explains ``significantly suppressed atmospheric gravity waves at locations of magnetic flux'' found by \cite{straus08}. The ubiquity of near-horizontal field in the low solar atmosphere discovered recently with \emph{Hinode} \citep{lites08} however raises the possibility that low frequency gravity waves may efficiently couple to Alfv\'en waves that continue to propagate vertically into the corona, contributing to the vast sea of waving field lines now known to exist there \citep{pontieu07,tomczyk07}. Although \citeauthor{tomczyk07} detect a peak in velocity power of these Alfv\'en oscillations at around 3.5 mHz, (presumably driven by the Sun's internal normal modes), the power spectrum continues to rise with decreasing frequency till at least 1 mHz, well inside the gravity wave regime at photospheric level. It is tempting to postulate that gravity waves may be the vector of this wave energy at low levels and that it may convert to Alfv\'en waves around the acoustic/Alfv\'enic equipartition level in highly inclined field regions. Further observational work, ideally at multiple heights, is warranted to more fully explore these possibilities.
}

It should be emphasised though that our analysis is entirely linear. Acoustic waves are likely to shock before reaching the upper chromosphere. However, Alfv\'en waves do not suffer this fate. Our models are also adiabatic, which is not a good representation of the chromosphere, {\hilite though the detections of \cite{straus08} suggest that {\glite atmospheric} radiative losses do not completely suppress gravity waves, at least at the {\glite photospheric} altitudes sampled by IBIS} {\glite and MDI}. The adiabatic assumption will be relaxed in future work.

\section*{ACKNOWLEDGMENT}
The authors thank Bernhard Fleck and Friedrich Schmitz for providing the tabulated model atmosphere used in this paper.

%%%%%%%%%%%%%%%%%%%%%%%%%%%%%%%%%%%%%%%%%%%%%%%%%%%%%

%%%%%%%%%%%%%%%%%%
%\clearpage
\appendix
\section{Dispersion Relation}   \label{A}

As explained in \cite{sc06}, a convenient starting point for deriving the dispersion relation in MHD is the Lagrangian density associated with the linearized MHD equations
\citep{goedpoedts}. A little manipulation using the equilibrium condition
$\grad p=\rho\g+\j\vcross\B$ yields the elegant Hermitian form
\begin{multline}
\calL = \half\thth \rho\, |\dot{\bxi}|^2 -\half\rho\, c^2 |\Div\bxi|^2 -
\half|\b|^2 
\\{}
-\half\thth\g\thth\vdot\left(\rho^{1/2}\bxi\,\Div\rho^{1/2}\bxi^* + 
\rho^{1/2}\bxi^*\,\Div\rho^{1/2}\bxi\right)  \\
{}-\quart\thth \j\thth\vdot\left(\bxi^*\vcross \b
+ \bxi\vcross \b^*\right)  
-\quart\thth \j\vcross\B \thth\vdot \left(\bxi\Div\bxi^*+\bxi^*\Div\bxi\right)
\, ,                                                         
\label{L}
\end{multline}
where $\bxi$ is the plasma displacement vector and $\dot{\bxi}$ its time
derivative, $\rho$ the density, $c$ the sound speed, $\B$ the magnetic field,
$\j=\Curl\B$ the current density, $\g=-g\,\e_z$ the gravitational acceleration,
and $\b = \Curl(\bxi\vcross\B)$ the magnetic field perturbation. The magnetic
permeability $\mu$ has been scaled to unity for simplicity. Variation of the
action $\int\calL\,dV\,dt$ with respect to $\bxi$ recovers the linear wave
equations. The last term in Equation (\ref{L}) vanishes for a force free
magnetic field, and the last two terms for a potential field. For the case at hand, a uniform field, both terms are dropped. We choose to adopt $\X=\rho^{1/2}\bxi$ as the dependent variable in the non-magnetic terms, but to remain with $\bxi$ in the magnetic term $\curl(\bxi\vcross\B)$.

Following the standard eikonal method \citep{wein}, we may then identify
$\grad\X\equiv \irom\,\k\X$, and in particular $\Div\X\equiv \irom\,\k\vdot\X$ in the
non-magnetic terms. On the other hand, $\curl(\bxi\vcross\B)\equiv
\irom\,\k\vcross(\X\vcross\a)$ is adopted in the magnetic energy term. Assuming the
background state is stationary, $\partial\X/\partial t\equiv -\irom\,\omega\X$. Then
\begin{equation}
\begin{split}
\calL &= \half \omega^2|\X|^2 - \half c^2 \left|\k\vdot\X - \frac{\irom\,Z}{2H}\right|^2 -
\half\left|\k\vcross(\X\vcross\a)\right|^2 \\
&\qquad\qquad{}+
\frac{\irom\,g}{2}\left(Z^*\,\k\vdot\X - Z\,\k\vdot\X^*\right) \\  
% &= \half(\omega^2-a^2\kpar^2)|\X|^2 -\half(a^2+c^2)|\k\vdot\X|^2
% + \half a\kpar\, \a\vdot\left(\X\X^*+\X^*\X\right)\vdot\k\\
% &\qquad\qquad{}+ \frac{i}{2}\left(g-\frac{c^2}{2H}\right)
% \left(Z^*\,\k\vdot\X - Z\,\k\vdot\X^*\right) - \half\frac{c^2}{4H^2}|Z|^2\\
&= \half(\omega^2-a^2\kpar^2)|\X|^2 -\half(a^2+c^2)|\k\vdot\X|^2\\
&\qquad\qquad{}
+\half\left(\m\vdot\X\,(\k\vdot\X)^* + (\m\vdot\X)^*\,\k\vdot\X \right)
- \half \omega_{\rm c}^2|Z|^2,
\end{split}                                                   
\label{Lf}
\end{equation}
where $Z$ is the vertical component of $\X$, $\e_z$ is the upward vertical unit
vector,  $\a=\rho^{-1/2}\B=a(\sin\theta\cos\phi\,\e_x+\sin\theta\sin\phi\,\e_y+
\cos\theta\,\e_z)$ is the Alfv\'en velocity (with $a=|\a|$ the Alfv\'en speed),
and $\kpar=\hat\a\vdot\k=k\cos\alpha$ is the field-aligned component of $\k$.
Here, we have found it convenient to define the complex vector
\begin{equation}
\m = (\a\vdot\k)\,\a -\irom\left(g-\frac{c^2}{2H}\right)\e_z\, .   \label{m}
\end{equation}

From the Lagrangian density, we may read off the Hermitian dispersion tensor
$\D$ defined by $\calL=\half \X^H\D\X$ in matrix notation, and consequently its
determinant, the dispersion function, which simplifies to
\begin{multline}
\calD=\det\D=
\omega^2 \omega_{\rm c}^2 a_y^2  k_{\rm h}^2 +(\omega^2-a^2\kpar^2)\times{}\\
\left[\omega^4-(a^2+c^2)\omega^2 k^2+a^2c^2k^2\kpar^2 \right.\\
\left.{}+ c^2N^2 k_{\rm h}^2
-(\omega^2-a_z^2k^2) \omega_{\rm c}^2\right],   \label{DF}
\end{multline}
where $a_z$ is the vertical component of the Alfv\'en velocity and $a_y$ is the
component perpendicular to the plane containing $\k$ and $\g$. The {\bv}
frequency $N$ is defined by $N^2=g/H-g^2/c^2$, $ \omega_{\rm c}=c/2H$ is the isothermal
acoustic cutoff frequency, and $ k_{\rm h}$ is the horizontal component of
the wavevector.

\end{document}